\documentstyle[12pt,fullpage]{article}
\newcommand{\alku}{\begin{equation}}
\newcommand{\loppu}{\end{equation}}
\newcommand{\puoli}{\frac{1}{2}}

\newcommand{\eg}{{\it e.g.,\ }}
\newcommand{\ie}{{\it i.e.,\ }}
\newcommand{\etal}{{\it et al.}}

\newcommand{\al}{\alpha}
\newcommand{\be}{\beta}

\newcommand{\th}{\theta}
\newcommand{\vek}{\vec{k}}
\newcommand{\vep}{\vec{p}}

\newcommand{\bfS}{{\bf S}}
\newcommand{\rarr}{\rightarrow}

\title{\ \ \ \ \ \ \ \ \ \ \ \ \ \ \ \ \ \ \ \ \ \ \ \ \ \
 \ \ \ \ \ \ \ \ \ \ \ \ \ \ \ \ \ \ {\normalsize CTP\ \# 2213}\\
  On Coarse-grained Entropy and Stimulated Emission in Curved
  Space-Time}

\author{Esko Keski-Vakkuri\thanks{Supported in part
by funds provided by the U.S. Department of Energy (DOE)
under contract $\sharp $DE-AC02-76ER03069.} \\
    {\small Center for Theoretical Physics}\\
    {\small Laboratory for Nuclear Science}\\
    {\small and Department of Physics}\\
    {\small Massachusetts Institute of Technology}\\
    {\small 77 Massachusetts Avenue, Cambridge MA 02139}}

\date{August, 1993}

\begin{document}

\maketitle

\begin{abstract}

We study the entropy generation and particle production in scalar
quantum field theory in expanding
spacetimes with many-particle mixed initial states. The recently
proposed coarse-grained entropy approach by Brandenberger \etal \ is
applied to systems which may have a non-zero
initial entropy. We find that although the particle production
is amplified as a result of boson statistics, the (coarse-grained)
entropy generation is {\em attenuated} when initial particles are present.
\end{abstract}

\newpage

One of the interesting features of quantized fields
in a curved spacetime \cite{1}
is that the concept of particles becomes very observer-dependent. For instance,
in an expanding Universe spontaneous particle creation can occur.
One defines generally a vacuum state such
that all inertial observers in the past region agree that the spacetime
looks empty of particles. As a result of the expansion of the Universe,
the above vacuum state looks full of particles using modes natural
to inertial observers in the far future region.
Stated differently, a no-particle
initial state can evolve to a many-particle state. However, since one
starts with a pure state, one ends with a pure state. Thus there must be
subtle correlations between the particles in the final state. In particular,
there is no entropy production in this process even if lots of particles
are produced.
But, it may be that some of these subtle correlations are very difficult
to detect and/or that they may be quite sensitive to interactions between
the produced particles. One may then contemplate a picture where
such information about the system is considered to be ``less relevant''
and either discarded altogether or some kind of ``statistical averaging''
procedure is applied to it. This way one can
try to associate a ``coarse-grained'' entropy to the
final state of the system, hopefully in as natural way as possible.
There has been a lot of work in this direction
by Hu, Kandrup and collaborators \cite{15}.

Recently, two different such approaches have been  proposed.
Brandenberger, Mukhanov and
Prokopec (BMP) discussed in \cite{2,3} a coarse-graining procedure based on
averaging over the so called squeeze angles which appear in the S-matrix of
particle production. On the other hand, Gasperini and Giovannini \cite{4},
together with Veneziano (GGV) \cite{45} related entropy generation
to an increased
dispersion of a superfluctuant operator.
Both groups were especially interested in the entropy generation related to
the production of gravitational waves and
density fluctuations in inflationary universe models.

In this work, our aim is to follow the BMP approach and study
the entropy generation starting not from an initial vacuum state with
zero entropy, but
allowing the system to be initially in some generic
many-particle (mixed) state
with non-zero entropy. If one starts with many bosons it is known \cite{5}
that the particle production will be amplified as a result of boson statistics,
as one would expect. So, in general one can ask whether the entropy generation
(in the coarse-grained sense) would also be amplified or not.
Indeed, as a consistency check it is necessary to investigate if definitions
of coarse-grained entropy will lead to a growing entropy even if initial state
is allowed to be an arbitrary many-particle state with initial entropy.
In \cite{4,45} the GGV entropy was shown to be growing at least
in certain classes
of initial states. Interestingly, it was found that their entropy
generation  did not
depend at all on the number of particles or entropy of the initial state.
Here we will attempt to investigate the BMP entropy in similar situations.
At least in the case of an initial density matrix where particles appear
as pairs of opposite momenta, and initial entropy
depends on the average occupation
number per mode, we can show that the BMP entropy grows, though
the entropy generation is {\em attenuated}. The BMP
entropy {\em does} depend on the
initial number of particles in a non-trivial way.
In the end we comment briefly
on the case of an initial thermal density matrix.

A scalar field in a $D$-dimensional curved spacetime is described by an action
\alku
   S = \int d^Dx \sqrt{-g} \ \puoli [g^{\mu \nu} \partial_{\mu} \phi
       \partial_{\nu} \phi - (m^2 + \xi R(x)) \phi^2 ],
\loppu
where $R(x)$ is the Ricci scalar curvature of the metric and $\xi$
is a coupling constant. Assume that
the metric depends explicitly on time and that it is asymptotically flat
in the far past and far future \cite{1}:
\alku
   g_{\mu \nu} (\vec{x},t) \rarr C_{\pm} \eta_{\mu \nu}
\loppu
as $t \rarr \pm \infty $.
In this case there are two natural ways to
quantize the field $\phi$ in Heisenberg
picture :
\begin{eqnarray}
  \phi = \sum_{\vec{k}} a^{in}_{\vec{k}} u^{in}_{\vec{k}} + h.c. \\
  \phi = \sum_{\vec{k}} a^{out}_{\vec{k}} u^{out}_{\vec{k}} + h.c. \,
\end{eqnarray}
where $u^{in}_{\vec{k}}$ looks like a plane wave in the far past,
\alku
  u^{in}_{\vec{k}} (\vec{x},t) \propto e^{-i\omega^{in}_{\vec{k}} t
                                        + i\vec{k} \cdot \vec{x}}
  \ ,\ t \rarr -\infty
\loppu
and $u^{out}_{\vec{k}}$ looks like a plane wave in the far future, respectively
\alku
 u^{out}_{\vec{k}} (\vec{x},t) \propto e^{-i\omega^{out}_{\vec{k}} t
                                        + i\vec{k} \cdot \vec{x}}
  \ ,\ t \rarr +\infty \ .
\loppu
These solutions can be related via a Bogoliubov transformation, which
can be given in
terms of annihilation/creation operators as
\alku
   a^{in}_{\vec{k}} = \al^*_{\vec{k} \vec{p}} a^{out}_{\vec{p}}
                      - \be^*_{\vec{k} \vec{p}} a^{\dagger out}_{\vec{p}} \ .
\loppu
This transformation is generated by a S-matrix
\alku
  a^{in}_{\vec{k}} = \bfS a^{out}_{\vec{k}} \bfS^{-1} \ ,
\loppu
which has the explicit form \cite{65}
\alku
 \bfS = \frac{1}{\sqrt{\det (\al_{\vec{k} \vec{p}})}}
     : \exp \{ \puoli [\al^{-1} \be^*]_{\vec{k} \vec{k}'}
                    a^{\dagger out}_{\vec{k}} a^{\dagger out}_{\vec{k}'}
           + [\al^{-1} - 1]_{\vec{k} \vec{k}'}
                a^{\dagger out}_{\vec{k}} a^{out}_{\vec{k}'}
              - \puoli [\al^{-1} \be ]_{\vec{k} \vec{k}'}
                   a^{out}_{\vec{k}} a^{out}_{\vec{k}'} \} :  \ \ .
\loppu
The factor $\frac{1}{\sqrt{\det (\al_{\vec{k} \vec{p}})}}$ is the in-out
vacuum amplitude. We use the convention of \cite{6} where the coefficients
$\al$ have taken to be real. The S-matrix is known to generate a unitary
transformation between the 'in' and 'out' representations if the
gravitational field has a compact support \cite{7}. For Robertson-Walker
type universes the in-out vacuum amplitude is zero and the 'in' and 'out'
representations are thus unitarily inequivalent.

The S-matrix relates the in- and out-vacuum states in the following way
\begin{eqnarray}
   \mid 0,in \rangle &=& \bfS \mid 0,out \rangle \nonumber \\
       &=& \frac{1}{\sqrt{\det (\al )}}\
           \exp \{ \puoli [\al^{-1} \be]_{\vec{k} \vec{k}'}
        a^{\dagger out}_{\vec{k}} a^{\dagger out}_{\vec{k}'} \}
            \mid 0,out \rangle \ .
\end{eqnarray}
This is the statement that an inertial observer in the far future region
sees the in-vacuum state as full of out-particles.
Similarly, the density matrix of the system expanded using in-modes
($\equiv \rho_i$) can be related to an expression
using out-modes ($\equiv \rho_f$)
as follows
\begin{eqnarray}
     \rho_i &=& \prod_{\vek} \sum^{\infty}_{n_{\vek}=0} f(n_{\vek})
       \mid n_{\vek},in\rangle \langle in,n_{\vek} \mid \\ \nonumber
       &=& \prod_{\vek} \sum_{n_{\vek}} \frac{f(n_{\vek})}{n_{\vek} !}
            \bfS a^{\dagger out}_{\vek} \bfS^{-1} \cdots
            \bfS a^{\dagger out}_{\vek} \bfS^{-1} \bfS \mid 0, out \rangle
            \langle out , 0 \mid \bfS^{-1} \bfS a^{out}_{\vek} \bfS^{-1} \cdots
            \bfS a^{out}_{\vek} \bfS^{-1} \\ \nonumber
      &=& \prod_{\vek} \sum^{\infty}_{n_{\vek}=0} f(n_{\vek})
       \bfS \mid n_{\vek},out\rangle \langle out,n_{\vek} \mid \bfS^{-1}
          \ \equiv \ \bfS \rho_f \bfS^{-1} \ .
\end{eqnarray}
Suppose now that the system is initially in an arbitrary many-particle state.
In this state the average occupation number per mode (using in-modes)
is given by
\alku
  \bar{n}^i_{\vec{k}} = \frac{1}{Tr \rho_i} Tr(\rho_i
       a^{\dagger in}_{\vec{k}} a^{in}_{\vec{k}} ) \ .
\loppu
In the far future region an inertial observer sees the average occupation
number per mode using out-modes as
\alku
\bar{n}^f_{\vec{k}} = \frac{1}{Tr \rho_f} Tr(\bfS \rho_f \bfS^{-1}
       a^{\dagger out}_{\vec{k}} a^{out}_{\vec{k}} )
\loppu
Using the cyclicity of the trace and the
properties of the Bogoliubov transformation one can derive
the relation between $\bar{n}^f_{\vec{k}}$ and $\bar{n}^i_{\vec{k}}$
to be
\alku
  \bar{n}^f_{\vec{k}} = \bar{n}^i_{\vec{k}} +
   \mid \be_{\vec{k} \vec{p}} \mid^2 (1 + 2\bar{n}^i_{\vec{p}}) \ .
\loppu
This is the formula for ``stimulated emission'' \cite{5}.
It tells us that even if the spontaneous creation of particles is weak,
$\mid \be_{\vec{k} \vec{p}} \mid^2 \ll 1$, the particle production
can become arbitrarily large,
\alku
  \Delta \bar{n}_{\vec{k}} \equiv \bar{n}^f_{\vec{k}} - \bar{n}^i_{\vec{k}}
       = \mid \be_{\vec{k} \vec{p}} \mid^2 (1 + 2\bar{n}^i_{\vec{p}})
       \rarr  {\it large}
\loppu
if the initial average occupation number per mode is arbitrarily large.
This amplification of particle production
is a result of the boson statistics of the particles. For fermions
the particle production would be attenuated \cite{8}.

Let us now discuss for notational simplicity metrics of the form
\alku
   ds^2 = dt^2 - a^2(t) d^2 \vec{x}^2 \ ,
\loppu
where $a(t)$ is a scale parameter of the universe. We again just require
that $a(t) \rightarrow a_{\pm}$ asymptotically as $t \rightarrow \pm \infty $.
As a result of the invariance under
spatial translations, the Bogoliubov coefficents can be written as
\alku \begin{array}{l}
  \al_{\vek \vep} = \al_{\vek} \delta_{\vek,\vep} \equiv
             \cosh r_{\vek}  \delta_{\vek,\vep} \\
  \be_{\vek \vep} = \be_{\vek} \delta_{\vek,-\vep}
                 \equiv e^{i\th_{\vek}} \sinh r_{\vek} \delta_{\vek,-\vep}\ .
\end{array} \loppu
The parameters $r_{\vek},\ \theta_{\vek}$ are called squeeze parameter
and squeeze angle in the Quantum Optics litterature, and the S-matrix is
called a two-mode squeeze operator \cite{9}.
If one starts with a vacuum state, the final
state (10) is called a squeezed vacuum.
In the initial vacuum case, if one expands
the corresponding $\bfS \rho_f \bfS^{-1}$ in the 'out' basis of
energy eigenstates,
one finds \cite{3} that the
off-diagonal components of $\bfS \rho_f \bfS^{-1}$
have an oscillatory dependence of the
angles $\theta_k $. In the BMP coarse-grained entropy approach it is
assumed that these angles represent irrelevant information about the
system (\eg in the sense that they would be very difficult to measure)
and they are therefore averaged over. After the averaging only the diagonal
elements of $\bfS \rho_f \bfS^{-1}$ survive and one
then defines a coarse-grained entropy
with the resulting reduced density matrix $\rho_{red}$
with the usual formula $S = -k_B Tr(\rho_{red} \log \rho_{red})$.
The result is \cite{2,3}
\begin{eqnarray}
   S^{f,0} \equiv \sum_{\vek} s^{f,0}_{\vek} &\equiv &
          k_B\sum_k (\cosh^2 r_{\vek} \log \cosh^2 r_{\vek}
             - \sinh^2 r_{\vek} \log \sinh^2 r_{\vek} ) \nonumber \\
               & = & k_B\sum_{\vek}
        [(\bar{n}^{f,0}_{\vek} +1) \log (\bar{n}^{f,0}_{\vek} +1)
                 - (\bar{n}^{f,0}_{\vek}) \log (\bar{n}^{f,0}_{\vek}) ]\ ,
\end{eqnarray}
where the notation $\bar{n}^{f,0}_{\vek}$ means the LHS of (14) in the case of
an initial vacuum state.
Now we turn to consider initial density matrices $\rho_i$ which can
describe generic
many-particle states with non-zero entropy.
Let us assume that the initial density matrix has the form
\alku
 \rho_i = \prod_{\vek , (k_z > 0)} \sum^{\infty}_{n=0}
  f_{\vek} (n) \mid n_{\vek}, n_{-\vek}, in \rangle
  \langle in, n_{\vek}, n_{-\vek} \mid \ \ ,
\loppu
where the coefficients $f(n_{\vek})$ are of
the form $f(n_{\vek}) = (\bar{n}^i_{\vek})^{n}/
(\bar{n}^i_{\vek} +1)^{n+1}$.
That is, we start with a many-particle state where particles
appear in pairs of opposite momenta, with an initial
average occupation number spectrum
$\bar{n}^i_{\vek} = \bar{n}^i_{-\vek}$
and with (ordinary) entropy given by $-k_B Tr(\rho_i \log \rho_i )$.
Writing the initial entropy in more explicit form, it is
\alku
  S^i \equiv \sum_{\vek , (k_z >0)} s^i_{\vek} = k_B\sum_{\vek ,(k_z >0)}
      [(\bar{n}^{i}_{\vek} +1) \log
 (\bar{n}^{i}_{\vek} +1) - (\bar{n}^{i}_{\vek}) \log (\bar{n}^{i}_{\vek}) ] \ .
\loppu
When we expand the resulting
final density matrix $\bfS \rho_f \bfS^{-1}$ (11) in 'out' energy
eigenstates, we find that also
in this case the off-diagonal elements have an oscillatory dependence
on the angles $\theta_{\vek}$. Therefore, following the BMP approach
 and averaging
over the angles, only the diagonal elements will survive. Thus the reduced
density matrix of the final state has the form
\alku
 \rho_{red} = \prod_{\vek , (k_z > 0)} \sum^{\infty}_{n=0}
  \bar{f}_{\vek} (n) \mid n_{\vek}, n_{-\vek}, out \rangle
  \langle out, n_{\vek}, n_{-\vek} \mid \ \ ,
\loppu
where $\bar{f}_{\vek} (n) = \langle n_{\vek}, n_{-\vek}, out \mid \bfS \rho_f
\bfS^{-1} \mid out, n_{\vek}, n_{-\vek} \rangle $. After some effort, one can
show that the coefficients have the form
\alku
   \bar{f}_{\vek} (n) =
    \frac{(\bar{n}^f_{\vek})^n}{(\bar{n}^f_{\vek} +1)^{n+1}} \ ,
\loppu
where $\bar{n}^f_{\vek}$ is the LHS of (14) with
the $\bar{n}^i_{\vek}$ of (19).
Thus, the final coarse-grained entropy is
\alku
   S^f \equiv \sum_{\vek ,(k_z >0)} s^f_{\vek} = k_B\sum_{\vek ,(k_z >0)}
                    [ (\bar{n}^{f}_{\vek} +1) \log
 (\bar{n}^{f}_{\vek} +1) - (\bar{n}^{f}_{\vek}) \log (\bar{n}^{f}_{\vek})] \ .
\loppu
The entropy depends only on the occupation number spectrum of particles in
the final state. This result is in agreement
with a similar formula given in \cite{2}
by a more heuristic argument to define entropy
of a statistical system
with a definite spectrum which
is valid both in and far out of thermodynamical equilibrium.
Let us now compare the entropy generation per mode in the initial vacuum
and initial many-particle cases. Denote
\begin{eqnarray}
  \Delta_0 s_{\vek} \equiv s^{f,0}_{\vek} - 0 \\
  \Delta s_{\vek} \equiv s^f_{\vek} - s^i_{\vek}  \ ,
\end{eqnarray}
where (24) applies to the former case and (25) to the latter case.
As a first consistency check, we find that
\alku
    \Delta s_{\vek} \geq 0
\loppu
so that the coarse-graining led to a growing entropy in our many-particle
case.
However,
as we compare (25) and (24) we find that
\alku
     \Delta s_{\vek} \leq \Delta_0 s_{\vek}  \ ;
\loppu
\ie the entropy generation is {\em attenuated} if one starts with many
particles present in the mode $\vek$. The equality holds iff
$\bar{n}^i_{\vek} = 0$. The difference
$\Delta s_{\vek} - \Delta_0 s_{\vek}$ depends symmetrically
on $\bar{n}^i_{\vek}$ and $\mid \be_{\vek} \mid^2$ and decreases monotonically
as either variable increases. Further, as both variables approach infinity,
\alku
     \Delta s_{\vek}  - \Delta_0 s_{\vek}
          \ \rarr \ \log 2 -1 \approx -0.31
\loppu
asymptotically.  This finite value is the maximum difference between
the generated entropies per mode.
Thus, unlike the GGV entropy,
the BMP entropy is {\em not} independent of the number of particles in
the initial state, but has some 'memory' about the initial occupation
numbers. Since entropy is a measure of loss of information,
it would appear that more information about the initial state of the system
is conveyed to the final coarse-grained state when stimulated emission
dominates the spontaneous particle production (since entropy generation
is attenuated).

Next case to be investigated would be an initial thermal density matrix
$$
\rho_i = \prod_{\vek} Z^{-1}_{\vek} \exp (-\beta \omega^{in}_{\vek}
a^{\dagger in}_{\vek} a^{in}_{\vek})\ .
$$
This situation is somewhat trickier
to deal with, for the following reason. Initially, the particles of opposite
momenta are uncorrelated. However, in the expansion of the universe the
particles are produced in pairs of opposite momenta. This induces correlations
between the opposite momenta in the final density matrix. It would have the
form
\alku
 \bfS \rho_f \bfS^{-1} = \prod_{\vek , (k_z > 0)} \sum^{\infty}_{n,m,n',m'=0}
  f_{\vek} (n,m;n',m') \mid n_{\vek}, m_{-\vek}, out \rangle
  \langle out, n'_{\vek}, m'_{-\vek} \mid \ \ ,
\loppu
where
\begin{eqnarray}
 & & f_{\vek} (n,m;n',m') = \nonumber \\
 & &  \langle n_{\vek}, m_{-\vek}, out \mid
  \frac{1}{Z_{\vek} Z_{-\vek}} e^{-\beta \omega^{in}_{\vek}
    [(\al_{\vek} a^{\dagger out}_{\vek} - \be_{\vek} a^{out}_{-\vek})
    (\al_{\vek} a^{out}_{\vek} - \be^*_{\vek} a^{\dagger out}_{-\vek}) +
     (\vek \mapsto -\vek )] }
   \mid out, n'_{\vek}, m'_{-\vek} \rangle \ .
\end{eqnarray}
Again, one can see that the $n\neq n'$ or $ m\neq m'$ components have an
oscillatory dependence of the squeeze angles and they vanish in the
coarse graining. However, the diagonal coefficients (those of the
reduced density matrix) will have a form $f_{\vek} (n,m)$ where
the dependence on $n$ and $m$ does not factorize. Hence the opposite
momenta have aquired correlations through the particle production and
the reduced density matrix is not of the same type as the initial one.
As advocated in \cite{45}, one would like to ignore the correlations
between different modes. In \cite{45} this problem was avoided
by replacing the two-mode squeeze operator by a one-mode squeeze operator,
but this does not appear entirely satisfactory, since
the particles are then not created in the correct way as pairs of
opposite momenta.
We would like to propose that the correlations between opposite momenta
could be ignored by proceeding to
define $\bar{f}_{\vek} (n) = \sum_{m}
f_{\vek} (n,m)$ and $\bar{f}_{-\vek} (m) = \sum_{n}
f_{\vek} (n,m)$. Then we would define the final reduced density matrix to
be
\alku
 \rho_{red} = \prod_{\vek , (k_z > 0)} \sum^{\infty}_{n,m=0}
  \bar{f}_{\vek} (n) \bar{f}_{-\vek} (m) \mid n_{\vek}, m_{-\vek}, out \rangle
  \langle out, n_{\vek}, m_{-\vek} \mid \ \ ,
\loppu
which is of the same type as the initial density matrix.
Now the final entropy would be given by
\alku
    S^f = -k_B \sum_{\vek} \sum_n \bar{f}_{\vek} (n)
          \log \bar{f}_{\vek} (n) \ .
\loppu
Unfortunately, at the present we do not have explicit formulas for
the coefficients $\bar{f}_{\vek} (n)$ or the final entropy. It would be very
interesting to see if the resulting expressions could depend on the final
average occupation number in the same fashion as in the earlier case.
We hope to be able to return to this question in the future.

Finally, let us clarify that even if we found a different result as
in the GGV approach, namely an entropy that depends on the number of
particles in the initial state, we are not arguing that it would mean
that the BMP
approach is 'better' than the GGV approach. As
stated in \cite{15}, it is good to have different definitions of
entropy, corresponding to loss of different information
about the system. Both BMP and GGV approaches have the virtue of
giving the correct average occupation number for particles
in the final state.
Otherwise the GGV approach appears to discard information
about the system a bit more generously,
since it leads to a greater growth of entropy.

\bigskip

{\large {\bf Acknowledgements}}

\bigskip

It is a pleasure to thank prof. S.D. Mathur for guidance and numerous
discussions. I am also grateful to profs. R. Brandenberger,
A.J. Niemi and S. Stenholm for helpful discussions.

\end{document}